# Persistent Dirac Fermion State on Bulk-like Si(111) Surface

Jian Chen[1], Wenbin Li[1], Baojie Feng[1], Peng Cheng[1], Jinglan Qiu[1], Lan Chen[1*] and Kehui Wu[1,2*]

[1] *Institute of Physics, Chinese Academy of Sciences, Beijing 100190, China*

[2] *Collaborative Innovation Center of Quantum Matter, Beijing 100871, China*

*Corresponding Authors. lchen@iphy.ac.cn (L. Chen) & khwu@iphy.ac.cn (K. H. Wu)

## Abstract

The "multilayer silicene" films were grown on Ag(111), with increasing thickness above 30 monolayers (ML). We found that the "multilayer silicene" is indeed a bulk Si(111) film. Such Si film on Ag(111) always exhibits a ($\sqrt{3}\times\sqrt{3}$)R30° honeycomb superstructure on surface. Delocalized surface state as well as linear energy-momentum dispersion was revealed by quasiparticle interference patterns (QPI) on the surface, which proves the existence of Dirac fermions state. Our results indicate that bulk silicon with diamond structure can also host Dirac fermions, which makes the system even more attractive for further applications compared with monolayer silicene.

Silicon is the basis of modern microelectronics industry. Unlike carbon which exhibits both $sp^2$ and $sp^3$ allotropes represented by graphite and diamond, silicon has only diamond-like structure with $sp^3$ bonding in nature. Recently, silicene, a single sheet of Si atoms arranged in honeycomb lattice with hybridized $sp^2/sp^3$ bonding, has been predicted [1, 2] and successfully fabricated [3-13]. Similar to graphene, the band structure of silicene hosts Dirac fermions and hence exotic properties for potential applications such as quantum spin Hall effect and spintronics devices [2]. However, recent studies revealed strong influence of the substrate on both the atomic and electronic structure of monolayer silicene, and that the Dirac state may no longer exist in monolayer silicene on metal substrates [4, 14-18]. This remains a serious challenge to any further research and application of silicene.

In this Letter, we overcome this problem by pointing out that Dirac electron state can exist on thick, bulk-like Si(111) film surface, provided that the surface is ($\sqrt{3}\times\sqrt{3}$)R30° reconstructed. Our study was motivated by previous reports of "multilayer silicene" film on Ag(111) [8, 19, 20]. We performed a comprehensive study on the "multilayer silicene" films grown on Ag(111), with increasing thickness above 30 monolayers (ML) by scanning tunneling microscopy (STM) and spectroscopy (STS). We found, unexpectedly, that the "multilayer silicene" is indeed a bulklike silicon film with diamond structure, in other words, a Si(111) film. Strikingly, the film always exhibits a ($\sqrt{3}\times\sqrt{3}$)R30° honeycomb superstructure on the surface, in contrast to the well-known 7×7 reconstruction on bulk Si(111) surface. More interestingly, quasiparticle interferences (QPI) patterns were observed for all film surfaces with different thicknesses, and linear energy-momentum dispersion has been deduced, which proves the persistence of Dirac fermions on such Si film surfaces. The origination of this Dirac fermion state is explained by the honeycomb Si lattice on the surface. Such a persistent Dirac state on bulk silicon surface is even more promising for spintronics device applications, since it is easy to obtain, substrate effects are avoided, and directly compatible with the silicon microelectronic industry.

The experiments were performed in an ultra-high-vacuum chamber (base pressure better than $1.0\times10^{-10}$ Torr) equipped with a home-made low-temperature STM. A single-crystal Ag(111) substrate was cleaned by standard sputtering-annealing procedure. Silicon was evaporated from a heated Si wafer with a deposition flux of about 0.3 monolayers per minute. The STM observations were performed at liquid nitrogen temperature (77 K) with chemically etched tungsten tips. All the STM data were recorded in the constant-current mode with the bias voltage $V$ applied to the tip. The differential conductance (d$I$/d$V$) maps were extracted from the lock-in signal by applying a modulation of 20 mV at 777 Hz to the tip bias.

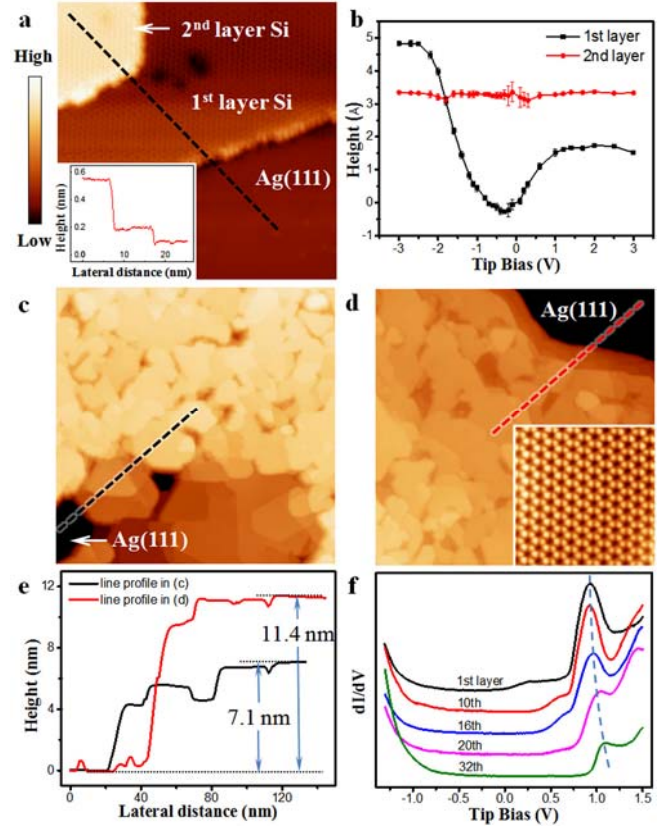

Fig.1. (a) STM image ($V_{tip}$ = 0.9 V, 20×20 nm$^2$) of the first two √3 layers of silicene film on Ag(111). The inset shows line profile along the black dash line in (a). (b) The curves indicate the apparent heights of each layer vary as a function of tip bias voltages. (c, d) STM images ($V_{tip}$ = -1.0 V, 200×200 nm$^2$) of multilayer silicene films with different thickness. The inset is high

resolution STM image ($V_{tip}$ = -0.5 V, 6×6 nm$^2$) on top terrace of film in (d). (e) The line profiles across the substrate to silicene films along the dash lines in (c) and (d), respectively. The thicknesses are indicated. (f) dI/dV curves obtained on surface of Si films with different thickness. The curves are vertically shifted for clarity.

A monolayer silicene film grown on Ag(111) surface exhibits a variety of different structural phases such as 4×4 [3, 21-24], √13×√13 [3, 4], √7×√7 [3, 24], 2√3×2√3 [23, 24] (with respect to Ag(111) surface lattice) and √3×√3 [3, 6](with respect to silicene 1×1). On the other hand, "multilayer silicene" films only exhibit √3×√3 honeycomb superstructure [25], as shown in Fig. 1(a). The line profile (as shown in inset) shows that the apparent heights of the first √3 layer and the second √3 layer at 0.9 V bias voltage are about 0.11 nm and 0.34 nm, respectively. We note, however, the apparent height of the first √3 layer varies significantly from 0 to 0.48 nm with the scan bias, as shown in Fig. 1(b). In contrast, the height of second √3 layer, 0.34 nm, is almost constant. This can be explained by the fact that the local density of states (LDOS) is different on the √3 layer surface and the Ag(111) substrate. On the other hand, the almost constant height of second √3 layer indicates the same LDOS for different √3 layers.

We prepared "multilayer silicene" films with thickness more than 30 layers (see Fig.1(c) and 1(d), the thickness can be determined by the line profiles). We observed, strikingly, that identical √3×√3 honeycomb superstructure persists on the surface up to the maximum thickness that we have obtained (inset in Fig. 1(d)). On the other hand, through careful analysis of the atomically resolved images near the step edges (Fig. 2(b) and 2(c)), we found that the stacking sequence of Si layers is strictly ABC stacking, for all different layer thickness. The observation of strict ABC stacking is beyond our expectation: If the neighboring Si layers interact with each other by weak van der Waals force just like graphite, one should AB stacking in most case. And due to the weak interlayer interaction, twisting between neighboring layers should be frequently observable. On the contrary, the stacking sequence of Si(111) layers in bulk silicon with diamond structure is strictly ABC stacking, as the strong covalent bonding between Si(111) layers effectively kills any mis-stacking. Therefore, we conclude that the so-called "multilayer silicene" is actually Si(111) film with √3×√3 honeycomb reconstruction on the surface, the atomic model being shown in Fig. 2(d). The above model is also supported by Raman measurements. It was reported that monolayer silicene exhibits Raman features different from bulk Si [26]. But when we measured the Raman spectroscopy on "multilayer silicene" film, the results showed only a 520 cm$^{-1}$ peak (Fig. S1), identical to bulk Si. While if the film structure is graphite-like, one should expected even stronger deviation of the spectrum from bulk Si due to the overlapping of signals from different graphite-like layers.

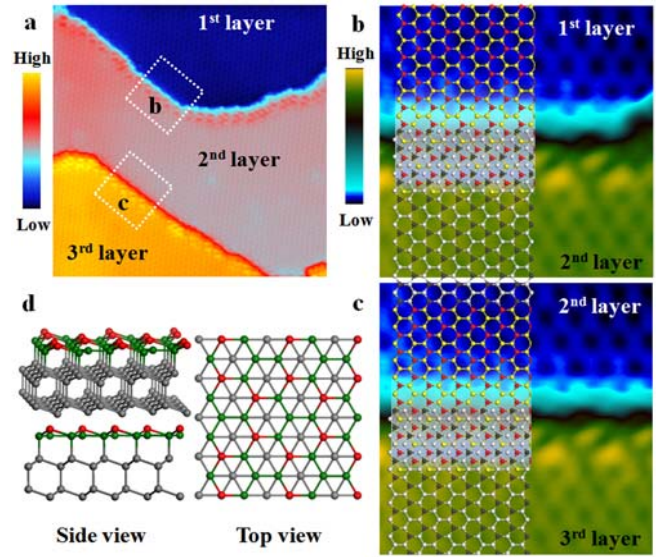

Fig. 2. (a) The STM image ($V_{tip}$ = 1.2 V, 25×25 nm$^2$) of Si film with thickness of three layers. (b) and (c) The high resolution STM images of area labeled by white squares in (a), respectively. The atomic model of the √3×√3 of silicene is superimposed, indicated that the stacking sequence of the neighbor Si layers is ABC stacking. (d) The side and top view of atomic model of Si film with (√3×√3)R30° reconstruction. The red and green balls represent silicon atoms with different buckling heights in first layer of Si film, respectively. The gray balls represent silicon atoms

below first layer.

Based on the above picture, a key question would be why a √3×√3 superstructure is formed on the surface, instead of the well-known 7×7, 5×5, or 2×1 on Si(111) [27, 28]. On the surface of bulk-terminated Si(111)1×1, there is one dangling bond per unit cell due to the symmetry breaking along Z direction. To lower the surface energy, one solution is to reduce the number of dangling bonds. In the case of Si(111)7×7, the number of dangling bonds in each unit cell is reduced from 49 to 19 [29]. However, the formation of 7×7 and 5×5 reconstructions of Si(111) surface requires temperature higher than 800°C due to the high energy barrier it has to overcome. In our experiments the sample temperature is lower than 300 °C during Si growth. If the substrate temperature is higher than 350 °C, the Si film would desorb completely. So the 7x7 reconstruction is not favored in our system. Another solution to lower the surface energy is to form a buckled surface by changing the bond angle and distance among surface atoms. The redistribution of electrons among the surface atoms may partially compensate the Si dangling bonds and stabilize the surface. In our present case, one of the two Si atoms in each honeycomb unit cell is buckled upward, forming the (√3×√3)R30° superstructure. In fact, many foreign atoms, such as Ag, Au, Cu, Al, Ga, Bi, B, etc., can form (√3×√3)R30° reconstruction on Si(111) surface [30-32], indicating that (√3×√3)R30° is a generally favorable configuration. If one treats the upper-buckled Si atoms as foreign atoms, the stabilization of silicon-√3×√3 superstructure can be at least partially understood. It is also notable that, the Si(111)(√3×√3)-Si had been reasonably used as a theoretical model to calculate the surface states of Si(111) surface [33]. Furthermore, the interface interactions between Ag(111) substrate and Si film may be another driving force to stabilize the (√3×√3)R30° reconstruction on Si(111) surface [34].

The most important issue is now whether such a (√3×√3)R30° superstructure induces a Dirac fermion state on the surface. For this STS measurements were performed to investigate the electronic structures of Si(111) film. Typical dI/dV curves obtained on surfaces with different thickness, as shown in Fig. 1(f), reveal similar features: a pronounced peak at positive bias 0.9V-1.1V and a DOS onset at negative bias 0.7-0.9V. While the film thickness increases, the positions of the peak and LDOS onset both shift slightly to the right. As we reported before [6, 35], the quasiparticles in monolayer silicene behave as Dirac fermions, similar with graphene. The identical dI/dV curves imply that the Dirac surface states may survive in Si(111) film with $\sqrt{3}\times\sqrt{3}$ honeycomb superstructure. In other word, the Dirac fermion is directly related to the unique surface state of $\sqrt{3}\times\sqrt{3}$ superstructure of Si(111), independent on the film thickness.

To further confirm this point we refer to differential conductance (dI/dV) maps. The quasiparticles in delocalized surface states scattered from the defects such as impurities or step edges will result in LDOS oscillations in differential conductance (dI/dV) maps. We have clearly observed such standing waves patterns in dI/dV maps on Si film surface with different thickness even higher than 30 layers. This indicates that the metallic surface state on our Si(111) surface is delocalized, and it should originate from the $\sqrt{3}\times\sqrt{3}$ superstructure on the surface, and not from the Ag(111) substrate. Because with a height >10nm, the LDOS of Ag(111) substrate should decay and be screened, and thus too small to be detectable on the Si surface. Note that although surface reconstructions of Si(111), such as 7×7 and 5×5 are metallic, the metallic surface state is localized on Si adatoms, and delocalized surface state has never been found. Only on some metal terminated Si(111) surface, for example Si(111) $\sqrt{3}\times\sqrt{3}$-Ag, the two dimensional free-electron-like surface state is found [36], in which case the surface states are mainly originated from s or p orbitals of Ag atoms [37].

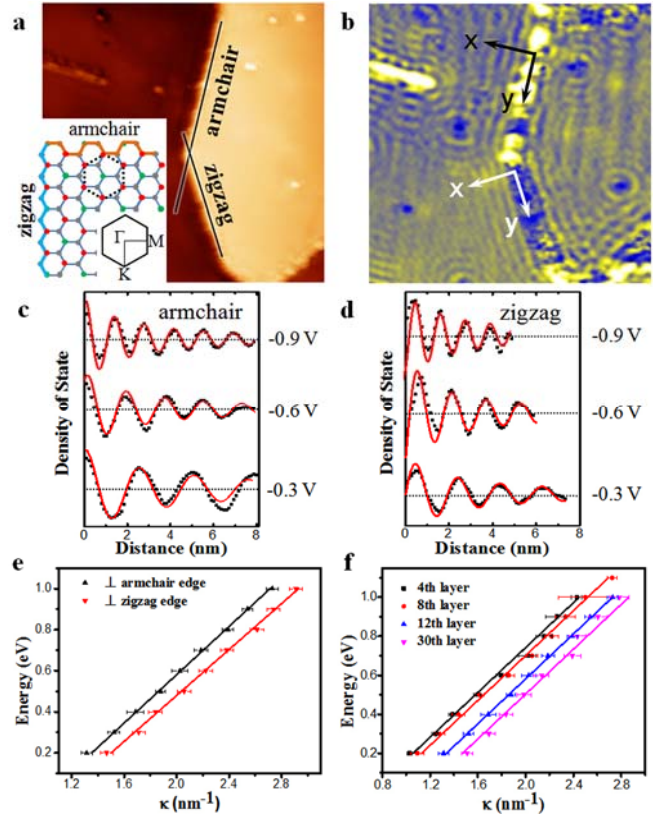

Fig. 3. (a) STM image ($V_{tip}$ = -0.4 V, 45×45 nm$^2$) obtained on top of Si film 20 layers, containing a island with both armchair and zigzag step edges. Inset: the atomic model of √3×√3 superstructure of Si. The black dash hexagon shows a honeycomb unit cell. The first Brillouin zone of the Si(111) 1×1 lattice is also shown. (b) dI/dV map ($V_{tip}$ = -0.4 V, 45×45 nm$^2$) of the same area as (a) showing obvious standing wave. (c, d) Line profiles of LDOS along the *x* axis for armchair and zigzag edges labeled in (b) at various energies, respectively. The back dash lines are experimental values, and the red lines are the lines fitting to the data. (e) Energy-momentum dispersions (E-k) determined from wave length of standing waves from armchair and zigzag edges in (c) and (d), respectively. The Energy-momentum dispersion obtained from Ag(111) surface state is also shown. (f) E-k curves (armchair edge) obtained on surface of Si films with different thickness on Ag(111) surface.

To quantitatively investigate the energy-momentum dispersion of the surface state, we focus on the standing waves around step edges. There are two types of step edges: zigzag and armchair, as exemplified in Fig. 3(a). The dI/dV map (shown in Fig.

3(b)) taken at same area as Fig. 3(a) shows obvious standing waves near both step edges. We plot the dI/dV intensity as a function of the distance from the step edges at various energies, and some examples are shown in Fig. 3(c)-(d). Here the direction normal to the step edge is defined as *x* and the direction parallel with the step edge is defined as *y*. The dI/dV signal has been averaged in *y* direction to maximize the signal-to-noise ratio. Clear oscillatory and decaying behavior of dI/dV signal along *x* axis is observed. The wavelength varies with the bias voltage, in other word, energy. We drew E($\kappa$) curves to deduce the energy-momentum dispersion relation with $2\kappa=|q|$ (q is scattering vector). Fig. 3(e) displays examples of the E($\kappa$) curves deduced from standing waves near armchair and zigzag step edges of a Si(111) film with twentieth layers, corresponding to dispersions at Γ-K and Γ-M directions of Brillouin zone (BZ), respectively. All the curves show clear linear dispersion, which is a strong evidence for the existence of Dirac fermion on Si(111) film with √3×√3 superstructure.

The linear dispersion of QPI patterns can be understood by supposing elastic scattering of quasiparticles within a constant energy contour (CEC) of the same Dirac cone (usually called intravalley scattering [38]). For dispersions at Γ-K and Γ-M directions, the two linear E-$\kappa$ curves are not exactly parallel due to the warping of the Dirac cone. At certain energy, $\kappa_A$ is slightly larger than $\kappa_Z$, which reveals anisotropic distribution of k at different directions (Γ-K and Γ-M). The slopes of curves give the Fermi velocity $V_{F-A} = (0.90 \pm 0.05) \times 10^6$ m/s and $V_{F-Z} = (0.83 \pm 0.05) \times 10^6$ m/s, respectively. In contrast, the dispersion curves measured from the standing waves on the clean Ag(111) surface (which is originating from the surface state of Ag(111)) shows a deviation from the Si film and a parabolic dispersion [39, 40], corresponding to conventional 2D free electrons. It is also a strong evidence that the delocalized surface state on Si(111) film is not from the Ag(111) substrate.

Fig. 3(f) shows the deduced linear energy-momentum dispersions of surface states for thickness about 4, 8, 12 and 31 layers. The slopes of these lines corresponding to Fermi velocity of electron are exactly same, whereas the energy position of Dirac point (DP), determined by the κ=0 energy intercept, is shifted downward from -0.4eV to -0.7eV, which coincides with the right shift of peak position in dI/dV curves shown in Fig. 1(f) well. There are two possibilities to explain the thickness dependent shift of DP. One is the band bending due to the charge transfer from silicon to Ag substrate. If it is true, the Fermi energy and DP of Si film will shift upward compared with those of free standing silicene. When the thickness of multilayer silicene increases, the Fermi level of top layer silicene will move back to its original position gradually. Another explanation is based on double barrier tunneling junction model [41]. One junction is between STM tip and surface, and another between Si film and Ag(111) substrate. The bias voltage applied between sample and tip will be divided onto two junctions. The magnitude of voltage drop in each junction should is proportional to the resistance of junction. When the thickness of Si film increases, the resistance of the junction between Si and Ag also increases. So the voltage drop between the STM tip and the Si film surface decreases. As a result, the measured position of DP will be lowered.

At last, let us discuss why the (√3×√3)R30° reconstruction can induce a surface state behaving as delocalized Dirac state. In the case of Si(111)7×7, there is a dangling bond in each of the 19 adatoms per unit cell [29]. The half-filled dangling bonds supports metallic surface states which is, however, localized because the distance between the dangling bonds is too long for their electron orbits to overlap with each other. In the case of (√3×√3)R30°, the dangling bonds in the surface Si atoms is only partially compensated by buckling, while the distance between them are short enough (one lattice constant). Therefore, dangling bonds will overlap to form the delocalized surface states.  Because the dangling bonds (upper buckled Si atoms) are arranged in a honeycomb lattice similar to graphene, a Dirac state will naturally result: The mapping of the sub-lattice degree of freedom to a pseudospin is represented by the Dirac equation for electrons bound to the lattice. Note that the existence of Dirac fermions in artificially constructed molecular graphene by CO molecules on Cu(111) surface has already been confirmed [42].

Our present work reveals that the "multilayer silicene" on Ag(111) is actually Si(111) film with (√3×√3)R30° honeycomb superstructure on the surface. We observed delocalized surface states on this particular surface, and the deduced linear energy-momentum dispersion suggests the surface state is Dirac state. We believe that the unique (√3×√3)R30° honeycomb structure determines these novel physical properties. The discovery of such a new Dirac system directly based on bulk Si paves the way to combining the revolutionary concept such as quantum computing with the current microelectronics industry based on Si.

**Acknowledgements:** This work was supported by the MOST of China (Grants No. 2013CBA01600, 2012CB921703, 2013CB921702), and the NSF of China (Grants No. 11334011，11322431, 11174344, 91121003).

**Supporting Information**

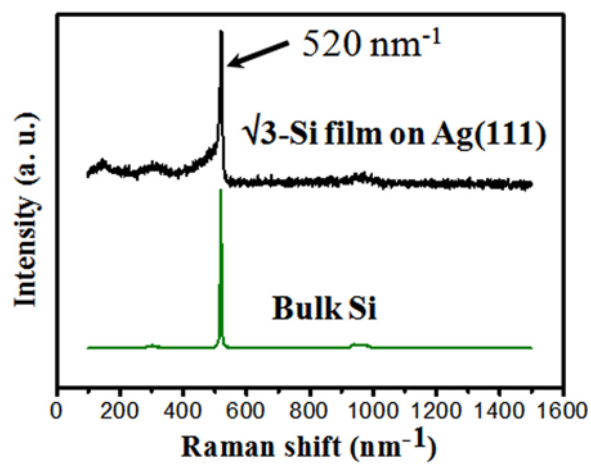

Fig. S1. Micro Raman spectra of a √3-Si film grown on Ag(111) surface, with thickness about 10 ML. The surface was confirmed by STM to have √3x√3 reconstruction before taking out of the UHV chamber for Raman measurements. Spectrum of a bulk Si wafer is included at the bottom as reference.